\PassOptionsToPackage{dvipsnames}{xcolor}
\documentclass[sigconf,nonacm]{acmart}

\usepackage{amsmath}
\usepackage{booktabs}
\usepackage{multirow}
\usepackage{multicol}
\usepackage[normalem]{ulem}
\useunder{\uline}{\ul}{}
\usepackage{appendix}
\usepackage{balance}
\usepackage{makecell}
\usepackage{color}
\usepackage[marginal]{footmisc}
\usepackage[linesnumbered,ruled,vlined, noend]{algorithm2e}
\usepackage{varwidth}
\usepackage{caption}
\usepackage{graphicx}
\usepackage{float} 
\usepackage{algorithmic}
\usepackage{marvosym}

\usepackage{enumitem}
\usepackage[dvipsnames]{xcolor}
\usepackage{subcaption}

\definecolor{bgcolor}{RGB}{242, 242, 242}
\setitemize[1]{topsep=0pt}

\newcommand{\stitle}[1]{\vspace{1ex} \noindent{\bf #1}}

\AtBeginDocument{%
  \providecommand\BibTeX{{%
    \normalfont B\kern-0.5em{\scshape i\kern-0.25em b}\kern-0.8em\TeX}}}

\begin{document}

\title{RuleAgent: Discovering Rules for Recommendation Denoising with Autonomous Language Agents}

\author{Zongwei Wang}
\email{zongwei@cqu.edu.cn}
\orcid{0000-0002-9774-4596}
\affiliation{%
  \institution{Chongqing University}
  \country{China}
}

\author{Min Gao}
\email{gaomin@cqu.edu.cn}
\authornote{Corresponding author}
\affiliation{%
  \institution{Chongqing University}
  \country{China}}

\author{Junliang Yu}
\email{jl.yu@uq.edu.au}
\affiliation{%
  \institution{The University of Queensland}
  \country{Australia}}

\author{Yupeng Hou}
\email{yphou@ucsd.edu}
\affiliation{%
  \institution{University of California San Diego}
  \country{United States}}

\author{Shazia Sadiq}
\email{shazia@eecs.uq.edu.au}
\affiliation{%
  \institution{The University of Queensland}
  \country{Australia}}

  \author{Hongzhi Yin}
\email{h.yin1@uq.edu.au}
\affiliation{%
  \institution{The University of Queensland}
  \country{Australia}}

\renewcommand{\shortauthors}{Zongwei Wang et al.}


\begin{abstract}
The implicit feedback (e.g., clicks) in real-world recommender systems is often prone to severe noise caused by unintentional interactions, such as misclicks or curiosity-driven behavior. A common approach to denoising this feedback is manually crafting rules based on observations of training loss patterns. However, this approach is labor-intensive and the resulting rules often lack generalization across diverse scenarios. To overcome these limitations, we introduce RuleAgent, a language agent-based framework which mimics real-world data experts to autonomously discover rules for recommendation denoising. Unlike the high-cost process of manual rule mining, RuleAgent offers rapid and dynamic rule discovery, ensuring adaptability to evolving data and varying scenarios. To achieve this, RuleAgent is equipped with tailored profile, memory, planning, and action modules and leverages reflection mechanisms to enhance its reasoning capabilities for rule discovery. Furthermore, to avoid the frequent retraining in rule discovery, we propose LossEraser—an unlearning strategy that streamlines training without compromising denoising performance. Experiments on benchmark datasets demonstrate that, compared with existing denoising methods, RuleAgent not only derives the optimal recommendation performance but also produces generalizable denoising rules, assisting researchers in efficient data cleaning.
\end{abstract}

\ccsdesc[500]{Information systems~Recommender systems}

\keywords{Recommendation, Agent, Denoising, Large Language Models}





\maketitle

\section{Introduction}
Recommender systems~\cite{57yin2019social,24wang2018minimax} have proven highly effective in domains like E-commerce~\cite{02yu2021self}, where they leverage historical user behavior data to uncover latent preferences and improve user experience. 
While explicit user feedback (e.g., ratings) is the most reliable choice for these system inputs, obtaining such feedback often requires active user engagement. As a result, implicit feedback (e.g., clicks) generated during user browsing is commonly used as an alternative~\cite{12DBLP:conf/sigir/Gao0HCZFZ22}. However, implicit feedback is often noisy~\cite{32wu2021ready}, as users may click on items out of curiosity or due to accidental misclicks~\cite{27bian2021denoising,28pan2013gbpr}. This noise exacerbates inaccuracies in estimating user preferences, underscoring the critical need for noise reduction strategies to improve recommendation accuracy~\cite{30wang2021implicit}.

\begin{figure}[t]
    \centering
    \begin{subfigure}[b]{0.4\textwidth}
        \centering
        \includegraphics[width=\textwidth]{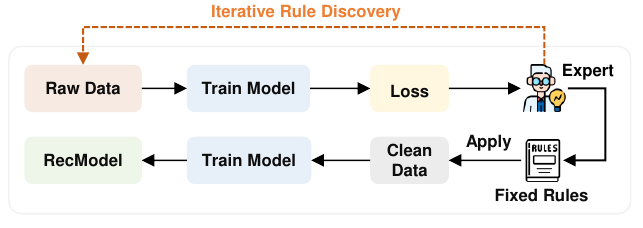}
        \caption{Expert-Guided Rule Discovery}
        \label{fig:image1}
    \end{subfigure}    
    \hfill
    \begin{subfigure}[b]{0.4\textwidth}
        \centering
        \includegraphics[width=\textwidth]{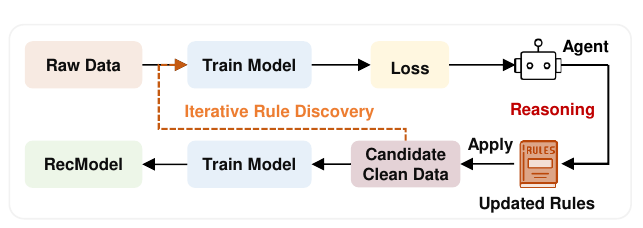}
        \caption{Agent-Driven Rule Discovery}
        \label{fig:image2}
    \end{subfigure}

    \caption{Expert-Guided vs. Agent-Driven Rule Discovery for Recommendation Denoising.}
    \label{introduction}
    \vspace{-1em}
\end{figure}

A common way to mitigate noise in implicit feedback involves utilizing auxiliary information, such as attribute features~\cite{44zhang2022neuro} or external data (e.g., social networks)~\cite{02yu2021self}. Unfortunately, such information is often limited, and the use of external data may raise privacy concerns~\cite{11wang2021denoising}, which complicates its practical application. In the absence of auxiliary information, numerous denoising approaches have emerged, including probabilistic-based frameworks~\cite{10wang2022learning}, adversarial training strategies~\cite{38he2018adversarial}, and causal inference-driven methods~\cite{93bonner2018causal}, etc. Among these, rule-based approaches have become the most widely adopted, relying on expert-defined criteria inferred from the dynamic patterns of loss values during training to estimate the noisiness of individual samples~\cite{11wang2021denoising,10wang2022learning}. For example, Wang et al. ~\cite{11wang2021denoising} found that a sample with high loss is likely to be noisy, while another study~\cite{10wang2022learning} demonstrated that noisy samples show large variations in loss across different models. As shown in Figure \ref{introduction} (a), these observations can be formulated into denoising rules that enable the identification and removal of noisy samples, allowing for model retraining to enhance both accuracy and robustness.

While rule-based methods have demonstrated notable effectiveness, they are inherently limited by several key limitations. Primarily, the formulation of denoising rules requires substantial input from domain experts, often relying on an iterative trial-and-error process. This not only demands considerable time and resources but also necessitates a deep understanding of the specific data context~\cite{12DBLP:conf/sigir/Gao0HCZFZ22,66he2024double}. Furthermore, once a denoising rule is established, its applicability to new contexts is far from guaranteed. When faced with new scenarios, the rule’s effectiveness must be re-evaluated, and if it proves ineffective, entirely new rules must be devised. This dependence on labor-intensive, expert-driven rule development, coupled with the limited transferability of these rules, severely limits their scalability and practical applicability.

To address the limitations inherent in current methods, we shift our focus to the recent rise of large language models (LLM) and LLM-powered agent-based applications~\cite{76wang2024survey,77zhao2023survey}. These studies leverage the powerful reasoning capabilities of LLM to create agents that can perform a variety of human-like tasks, such as daily activities~\cite{75park2023generative}, debates~\cite{73chan2023chateval}, and gaming~\cite{74chen2023agentverse}. Inspired by these innovations, we naturally consider whether LLM-powered agents could take on the role of data experts in uncovering denoising rules. As shown in Figure \ref{introduction} (b), if an agent can autonomously interact with data and models, iteratively evaluating hypotheses derived from data observations and model feedback, it could potentially discover denoising rules independently. In contrast to the labor-intensive and time-consuming process of manually identifying these rules, such an autonomous agent-based approach promises efficiency improvements. Additionally, the denoising rules discovered by the agent are dynamically aligned with the current training data and model, ensuring greater adaptability in handling diverse scenarios.

In this paper, we introduce \textbf{RuleAgent}, an autonomous language agent-based framework developed to discover rules for denoising recommendations. To suit the task of uncovering denoising rules, we have customized RuleAgent with four key modules: the \textit{Profile Module}, which defines the agent's role and the denoising task; the \textit{Memory Module}, which stores relevant information throughout the rule discovery process, such as historical behavior states, summaries of denoising rules, and the confidence scores of individual samples; the \textit{Planning Module}, which determines the next action of the agent from multiple perspectives; and the \textit{Action Module}, which defines the agent’s available action space, including assessing sample noise, refining denoising rules, and manipulating the model. 

To further enhance its autonomous analysis capabilities, these modules are integrated with two \textit{reflection} mechanisms: one for refining denoising rules within a hierarchical structure based on ongoing discoveries, and another for adjusting confidence levels regarding noisy samples. Additionally, recognizing that the agent's continuous actions require frequent re-training of the recommendation model, which consumes significant time, we adopt the concept of unlearning~\cite{80chen2022recommendation,81li2024ultrare} and propose a tailored strategy named \textit{LossEraser}. This strategy reclassifies agent-identified noisy samples as negative examples instead of unobserved ones, enabling the model to effectively ``erase'' the influence of these samples that were previously optimized as positive examples during earlier training. Consequently, the need for full re-training of RuleAgent is elegantly mitigated, maintaining efficiency without compromising performance. Our contributions are summarized as follows:
\begin{itemize}[leftmargin=12pt]
    \item We investigate the potential of language agents for rule discovery to denoise implicit feedback, marking the first application of language agents for data denoising in recommender systems.
    \item We propose RuleAgent, equipped with specialized modules and mechanisms to enhance its reasoning capabilities for rule discovery and a tailored unlearning strategy to enhance its efficiency.
    \item Experimental results on three datasets demonstrate that our proposed approach outperforms current state-of-the-art denoising recommendation models.
\end{itemize}


\section{Preliminary}
\subsection{Implicit Feedback Based Recommendation}
\stitle{General Recommendation.} 
Given user-item interaction data $\mathcal D=\{u,  i,  r_{u,i} | u \in \mathcal{U},i \in \mathcal{I}\}$, where $r_{u,i}=\{0,1\}$ indicates whether user $u$ has interacted with item $i$. Here, $\mathcal U\in\mathbb{R}^{|\mathcal U|}$ denotes the set of users, $\mathcal I\in\mathbb{R}^{|\mathcal I|}$ denotes the set of items. Generally, recommendation algorithms process the interaction data $\mathcal D$ to derive latent representations for users and items, $\mathbf{Z}_{\mathcal U}\in\mathbb R^{|\mathcal U|\times d}$ ($d$ is the dimension of representations) and $\mathbf{Z}_{\mathcal I}\in\mathbb R^{|\mathcal I|\times d}$, and a model $f$ with parameters $\theta_{f}$ to make predictions. The training of the recommendation model is formally defined by the optimization problem:

\begin{equation}
	\setlength{\abovedisplayskip}{-5pt}
	\setlength{\belowdisplayskip}{-1pt}
	\label{equation0}
	\begin{split}
		\quad \theta_{f}^{*}=\mathop{\arg\min}\limits_{\theta_{f}}  \mathcal{L}_{rec}(\mathcal{D}),
	\end{split}
\end{equation}
where $\theta_{f}^{*}$ is the optimal parameters of $f$, and $\mathcal{L}_{rec}$ is the recommendation loss herein instantiated with the BPR loss~\cite{28pan2013gbpr}:
\begin{equation}
	\label{equation1}
	\begin{split}
		\mathcal{L}_{rec} = \underset{(u, i, j) \in \mathcal{D}} {\mathbb{E}} -log( \sigma(f({\mathbf{z}_{u}}, \mathbf{z}_{i})-f({\mathbf{z}_{u}},\mathbf{z}_{j}))),
	\end{split}
\end{equation}
where the tuple $(u, i, j)$ consists of a user $u$, a positive sample item $i$ that $u$ has interacted with, and a negative sample item $j$ that $u$ has not interacted with. $\sigma$ is the sigmoid function.

\stitle{Recommendation Denoising.}
In the absence of auxiliary information, a prevalent approach to denoising implicit feedback involves analyzing the corresponding loss values to identify and eliminate noisy samples, thereby constructing a cleaner dataset~\cite{11wang2021denoising,92qin2021world}. Existing denoising methods typically rely on manually crafted denoising rules, developed through labor-intensive processes, to filter out noisy samples and derive a model trained with a cleaner dataset. This process can be formulated as follows:
\begin{equation}
\label{equation2}
\mathop{\arg\min}\limits_{\theta_{f}}\mathcal{L}_{rec}(\mathcal{D}_{c}), \quad s.t., \mathcal{D}_{c}=\mathcal{R}(\mathcal{L}_{rec}(\mathcal{D})),
\end{equation}
where $\mathcal{D}_{c}$ represents the rule-refined clean interaction dataset, and $\mathcal{R}$ refers to a rule function derived from the analysis of loss value, which guides the construction of $\mathcal{D}_{c}$.

From Equation \ref{equation2}, it is evident that the effectiveness of the process heavily depends on the $\mathcal{R}$ function. However, this $\mathcal{R}$ is not always straightforward due to the labor-intensive process of rule discovery and its limited transferability. To overcome these challenges, we explore the use of autonomous LLM-powered agents in addressing the recommendation denoising problem.


\subsection{LLM-Powered Autonomous Agents}
LLMs have recently exhibited significant potential for achieving human-level intelligence, sparking a growing interest in the development of autonomous agents. The architecture of these agents is generally organized into four primary modules: the \textbf{profile module}~\cite{84hong2023metagpt}, the \textbf{memory module}~\cite{78tu2023chatlog}, the \textbf{planning module}~\cite{79miao2023selfcheck,87wei2022chain}, and the \textbf{action module}~\cite{89wang2023voyager}. These modules are inherently interconnected, working together to enable the agent's functionality. The profile module defines the agent’s role and task, serving as the foundation for the memory module, which operates in a dynamic environment by storing past behaviors and providing context for future planning. The planning module acts as the decision maker, analyzing stored information and determining the next action. The action module executes these planning actions, carrying out tasks aligned with the agent's objectives. A key aspect of this architecture is the use of \textbf{reflection} mechanism~\cite{90brohan2023can,91zhu2023ghost}, which enhances the agent's autonomy by refining the profile and memory modules, facilitating memory updates, and improving the planning-making process for future actions.

In our approach, we treat the LLM-powered autonomous agent as a denoising expert. The agent autonomously interacts with the recommendation model, using reflection and corresponding actions to update its memory. Through this iterative process, the agent continuously refines the denoising rules, identifying noisy samples and constructing cleaner datasets. This process is defined as:
\begin{equation}
\label{equation3}
\mathop{\arg\min}\limits_{\theta_{f}}\mathcal{L}_{rec}(\mathcal{D}_{c}), \quad s.t., \mathcal{D}_{c}=\text{Agent}(f, \mathcal{D}),
\end{equation}
where $\text{Agent}$ represents the autonomous agent's multi-round interactions with the recommendation model $f$ and data $\mathcal{D}$. Ultimately, this iterative process enables the agent to identify a cleaner interaction dataset $\mathcal{D}_{c}$ through continuous operations. Compared with the original $\mathcal{R}$ function, the $\text{Agent}$ eliminates the need for complex rule mining and design prior to recommendation model training. Furthermore, it ensures that the function is no longer fixed but instead dynamically adapts to the patterns within the current dataset. 

\section{Module Design of RuleAgent}
\begin{figure*}[t]
  \centering
  \includegraphics[width=1\linewidth]{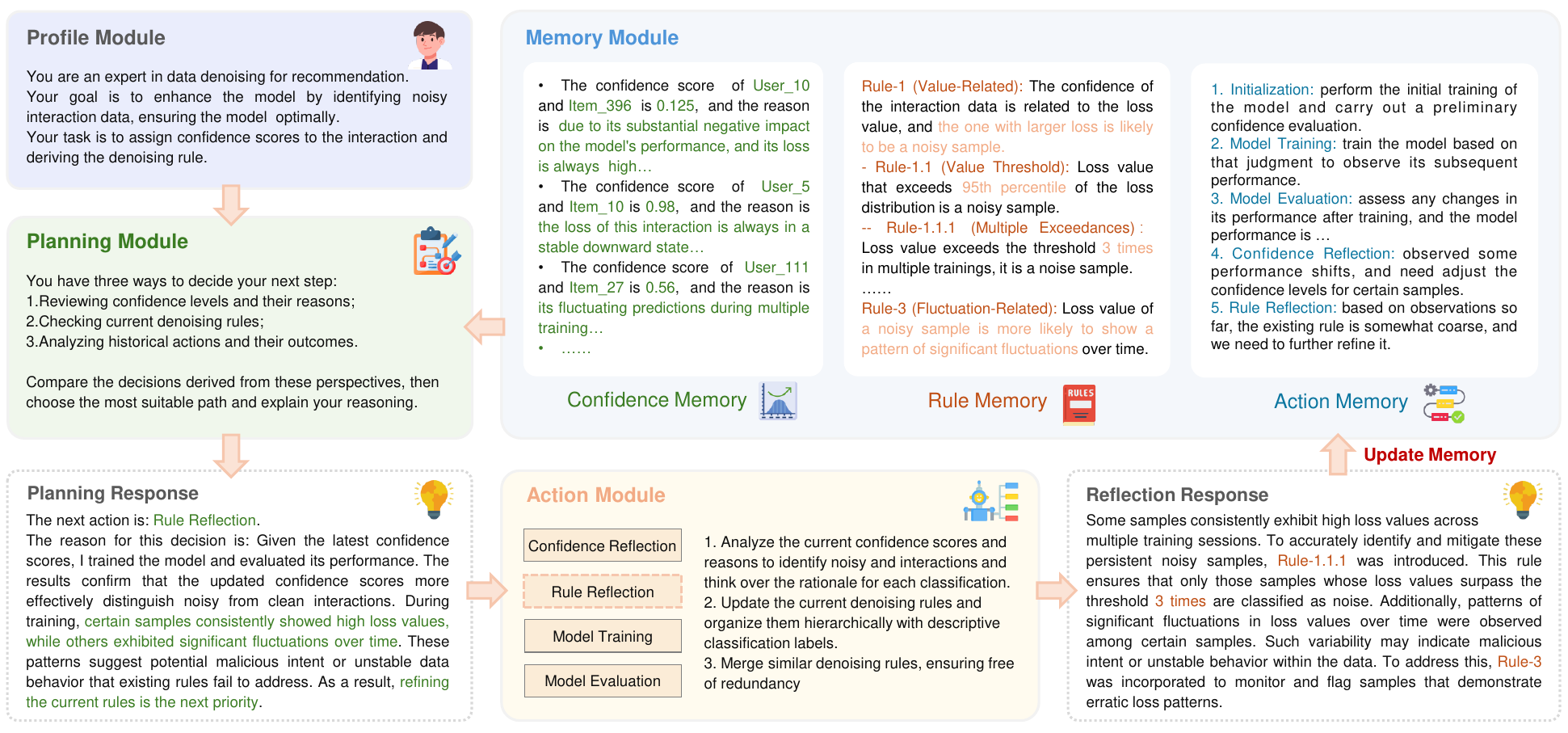}
  \caption{The Workflow of RuleAgent. It integrates Profile, Memory, Planning, and Action modules, with reflection mechanisms for updating Rule and Action Memories. Note that the descriptive text within each module in this figure is abbreviated.}
\label{method}
\end{figure*}

In this section, we introduce \textbf{RuleAgent}, an autonomous language agent-based framework developed to discover rules for recommendation denoising. The primary motivation behind RuleAgent is to mimic a real-world data expert, who explores and experiments iteratively to uncover denoising rules, applies these rules to identify noisy samples, and ultimately improves recommendation performance. The overall framework of RuleAgent is illustrated in Figure \ref{method}, which is designed around several key components: a profile module that defines the denoising task and establishes the agent’s role as a denoising expert; three updatable memory modules that store past states and learned denoising contexts; a planning module that dynamically makes a multi-path planning for determining the next action; and four action modules that specify the types of actions the agent can execute.

\subsection{Profile Module} To enable the agent to understand the context of the recommendation denoising scenario, we explicitly define its role, primary objectives, and the fundamental process of the task through the profile module. To demonstrate this approach, an example prompt serves as the content of the profile module, as shown below:\vspace{3pt}\\
\setlength{\fboxsep}{5pt}
\colorbox{bgcolor}{\begin{varwidth}{\dimexpr\linewidth-2\fboxsep\relax}
\small
\emph{\textcolor{purple}{You are an expert in data denoising.} \\ \textbf{Your goal} is to enhance the recommendation model by filtering out noisy interaction data, ensuring the model performs optimally. \textbf{Your task} involves assigning confidence scores to the interactions and deriving the appropriate denoising rules. Interactions with higher confidence will be kept for model training, while those with lower confidence will be discarded.}
\end{varwidth}}

\subsection{Memory Module}
\label{Memory Module}

\subsubsection{Confidence Memory.} To enable the agent to retain its evaluations of the confidence score for each sample, we design a unified sample-level memory to store the noise confidence scores. With the support of this confidence memory, the agent can directly retrieve and access each sample's noise confidence without requiring additional LLM inference. An example of a noise confidence score and its corresponding explanation for a single sample in the confidence memory is shown below:\vspace{3pt}\\
\setlength{\fboxsep}{5pt}
\colorbox{bgcolor}{\begin{varwidth}{\dimexpr\linewidth-2\fboxsep\relax}
\small
\emph{The noise confidence level for the interaction between \textcolor{olive}{<UserIndex>} and \textcolor{teal}{<ItemIndex>} is evaluated as \textcolor{purple}{<Score>}. The reason is \textcolor{purple}{<Reason>}.}
\end{varwidth}}

\subsubsection{Rule Memory.} The agent is expected to continuously summarize and refine the insights gained from the data and model training. To achieve this, we design the rule memory to store the denoising rules discovered and refined by the agent. The rule memory acts as the foundational principle guiding the agent in evaluating the noise confidence level of each sample. Furthermore, all stored rules are organized hierarchically, which promotes efficient management and analysis, enabling the seamless addition, removal, and merging of rules. The examples of rules are presented in Figure \ref{method}.

\subsubsection{Action Memory.} Additionally, we also design an action memory to record the agent's previous actions along with the corresponding observed feedback. This memory serves as a key resource for guiding the agent's planning-making in subsequent actions. An example of the action memory is presented below:\vspace{3pt}\\
\setlength{\fboxsep}{5pt}
\colorbox{bgcolor}{\begin{varwidth}{\dimexpr\linewidth-2\fboxsep\relax}
\small
\emph{The \textcolor{purple}{<i-th>} action is chosen as \textcolor{purple}{<Action>}, and the reason is \textcolor{purple}{<Reason>}.}
\end{varwidth}}

\subsection{Planning Module} The planning module serves as a crucial bridge between preceding and subsequent modules. Its primary function is to analyze information from both the profile module and the memory module to determine the next appropriate action. In RuleAgent, we have developed a multi-path planning method to ensure that the agent makes suitable decisions by considering confidence scores, denoising rules, and historical actions. More details are provided in Section \ref{planning-making}.

\subsection{Action Module} To ensure alignment with real-world rule discovery workflows, the action modules have been designed as simplified adaptations that handle four key tasks: confidence reflection, rule reflection, model training, and model evaluation. In the confidence reflection process, the agent employs a sample-level \textit{confidence reflection} mechanism, enabling it to autonomously update the confidence scores of individual samples. During the rule reflection process, we design a hierarchical \textit{rule reflection} mechanism for the agent to facilitate more refined and easily scalable rule discovery. The training and evaluating model steps involve the agent utilizing the memory to filter samples and form a new candidate clean dataset, which is then used to update the parameters of the recommendation model and monitor its performance. These steps serve to assess whether the current confidence scores and rules are reasonable and effective.

\section{Autonomous Workflow of RuleAgent}
In this section, we will thoroughly discuss the process of the agent's autonomous operation to elucidate how modules of RuleAgent interact with one another. 
\subsection{Initialization}
Before agent autonomously operating, it is essential to initialize each memory module. The rule memory is initialized with the ``Rule-1'' shown in Figure \ref{method}. For the confidence memory, we first perform a complete training cycle of the recommendation model, using the initial denoising rule and the loss values observed during training to assign noise confidence scores to each sample. Furthermore, the action memory records the first action as ``Initialization''.

\subsection{Multi-Path Planning}
\label{planning-making} Building on the initialized profile and memory modules, the agent autonomously adopts a multi-path planning approach, leveraging the information contained in these modules. Drawing on the principles of self-consistency~\cite{94wang2022self}, the planning module enables the agent to separately consider information from three distinct memory sources to gain situational awareness. By analyzing each memory source, the agent subsequently selects the most appropriate action from the defined action space. This multi-path planning approach can reduce potential biases and improve decision-making accuracy, and the process can be formulated as follows:
\begin{equation}
\label{equation7}
\begin{aligned}
M^{A} &\leftarrow {Planning}(M^{C}, M^{R}, M^{A}),
\end{aligned}
\end{equation}
where ${Planning}$ represents the multi-path planning process. It receives the current confidence memory $M^{C}$, rule memory $M^{R}$, and existing action memory $M^{A}$ as inputs. ${Planning}$ processes these inputs to generate a new action along with its corresponding reasoning, which are then stored in $M^{A}$. Below, we illustrate a simplified version of the multi-path planning system prompt:\vspace{3pt}\\
\setlength{\fboxsep}{5pt}
\colorbox{bgcolor}{
  \begin{varwidth}{\dimexpr\linewidth-2\fboxsep\relax}
  \small
  \emph{
  \textbf{System Prompt}: You have three potential planning paths for deciding the next step:\\
    1. Current Confidence-Based Planning: Examine the confidence levels and their associated
    reasons stored in \textcolor{purple}{<$M^{C}$>}. Determine the most suitable
    next step and explain your reasoning;\\
    2. Current Rule-Based Planning: Refer to the
    rules stored in \textcolor{purple}{<$M^{R}$>}. Decide on the most appropriate next step and provide the rationale;\\
    3. Historical Action-Based Planning: Analyze \textcolor{purple}{<$M^{A}$>}. Use these
    insights to determine the most suitable next action;\\
  Compare the decisions with three planning paths, then decide on the most suitable next step and explain your reasoning.
  }
  \end{varwidth}} 


\subsection{Action-Execution}
Once the planning module determines the next action, the agent performs the task within the action module. In the following, we provide a detailed explanation of the task for each action.

\subsubsection{Confidence Reflection.} The agent refines the sample confidence scores through the \textit{confidence reflection} mechanism, denoted as ${Reflection}^{C}$, when it identifies that the current scores assigned to the sampled data are inaccurate or insufficient. Specifically, the agent reviews the existing scores and their corresponding explanations stored in the confidence memory $M^{C}$. It then updates these values for each user-item interaction by utilizing the denoising rules from the rule memory $M^{R}$, alongside the historical loss values $\mathcal{L}_{u,i}$ obtained during the model training process. The process of updating a single sample's confidence can be formulated as follows:
\begin{equation}
\begin{aligned}
\label{equation8}
M^{C}_{u,i} \leftarrow {Reflection}^{C}(M^{C}_{u,i}, M^{R}, \mathcal{L}_{u,i}),
\end{aligned}
\end{equation}
where $M^{C}_{u,i}$ denotes the content in the confidence memory $M^{C}$ corresponding to the interaction between user $u$ and item $i$, and $\mathcal{L}_{u,i}$ represents the historical loss values associated with this interaction.
The brief system prompt of ${Reflection}^{C}$ is presented below:\vspace{3pt}\\
\setlength{\fboxsep}{5pt}
\colorbox{bgcolor}{
  \begin{varwidth}{\dimexpr\linewidth-2\fboxsep\relax}
  \small
  \emph{
    \textbf{System Prompt}: According to the rule \textcolor{purple}{<$M^{R}$>}, analyze the historical loss \textcolor{purple}{<$\mathcal{L}_{u,i}$>} and confidence score \textcolor{purple}{<$M^{C}_{u,i}$>}  of \textcolor{olive}{<UserIndex>} and \textcolor{teal}{<ItemIndex>}, and then update the \textcolor{purple}{<Score>} and explain the \textcolor{purple}{<Reason>}.
    }
  \end{varwidth}}

\subsubsection{Rule Reflection.}
The agent refines denoising rules using the \textit{rule reflection} mechanism, denoted as ${Reflection}^{R}$, whenever the current rule is identified as incorrect or suboptimal. During this process, the agent rethinks the existing denoising rules stored in the rule memory and incorporates information from the confidence memory. To address input token limitations, 1,000 interactions are randomly sampled for each \textit{rule reflection} process. If not stressed, this sampling method is consistently applied to retrieve confidence memory throughout the entire RuleAgent workflow. The process is formulated as follows:
\begin{equation}
\begin{aligned}
\label{equation9}
M^{R} \leftarrow {Reflection}^{R}(M^{C},M^{R}).
\end{aligned}
\end{equation}
The brief system prompt of ${Reflection}^{R}$ is presented below:\vspace{3pt}\\
\setlength{\fboxsep}{5pt}
\colorbox{bgcolor}{
  \begin{varwidth}{\dimexpr\linewidth-2\fboxsep\relax}
  \small
  \emph{
    \textbf{System Prompt}: Analyze and update denoising rules following these steps:\\
    1. Analyze current confidence scores and reasons in \textcolor{purple}{<$M^{C}$>}, and summarize why certain interactions are considered noisy and which are malicious.\\
    2. Compare the current denoising rule in \textcolor{purple}{<$M^{R}$>}. Then, organize the rules hierarchically with descriptive classification labels to ensure they are clear, concise, and actionable, such as "Rule-1(Value-Related)", "Rule-1.1(Label)", "Rule-1.1.1(Label)", ..., "Rule-N(Label)", "Rule-N.1(Label)", etc..\\
    3. Merge similar rules, ensuring the rules are free of redundancy.\\
    4. Output the updated denoising rules as \color{RoyalBlue}{<New Rules>}.}
\end{varwidth}}

\subsubsection{Model Training.}
When the agent believes that the memory has been sufficiently updated, it attempts to leverage the knowledge within the memory to construct a new clean dataset for retraining the recommendation model, aiming to achieve improved model parameters. However, retraining the recommendation model from scratch can be time-consuming. To overcome this issue, we introduce an efficient training method inspired by unlearning~\cite{80chen2022recommendation}, called \textit{LossEraser}. The core idea of LossEraser is to enable the model to efficiently ``forget'' the influence of noisy data. Specifically, we treat discarded samples as negative samples during optimization, thereby reversing the previously generated gradients when these samples were treated as positive during training. The LossEraser loss function is defined as follows: 
\begin{equation}
	\label{equation1}
	\begin{split}
        \mathcal{L}_{eraser} = \underset{(u, i, n) \in \mathcal{D}_{c}} {\mathbb{E}} -\log(\sigma(f(\mathbf{z}_{u},\mathbf{z}_{i})-  f(\mathbf{z}_{u},\mathbf{z}_{n}))),
	\end{split}
\end{equation}
where the tuple $(u, i, n)$ consists of a user $u$, a positive sample item $i$ that $u$ has interacted with, and an item $n$ that $u$ has also interacted with but is considered a noisy sample.

To achieve a stable and fine-grained unlearning process, we incorporate two key enhancements into the LossEraser method:
(1) \textbf{Progressive Reversal}: We adopt a time-dependent scaling factor, denoted as $\alpha_t$, which controls the extent of gradient reversal throughout the training process. Initially, $\alpha_t$ is set to a small value to avoid drastic changes to the model parameters. As training progresses, $\alpha_t$ gradually increases, enabling the model to progressively "forget" the influence of noisy data in a controlled manner. (2) \textbf{Adaptive Reversal}: Recognizing that not all noisy samples equally degrade model performance, we define a dynamic weight factor $w_{u,n} \in [0, 1]$ based on the confidence scores stored in memory. This weight factor modulates the strength of gradient reversal based on the severity of the sample's impact. Noisy samples with higher weights exert a stronger influence on the unlearning process, effectively "forgetting" more detrimental samples, while those with lower weights contribute less. Consequently, the updated loss function is defined as follows:
\begin{equation}
\begin{aligned}
\mathcal{L}_{eraser} = \underset{(u, i, n) \in \mathcal{D}_{c}} {\mathbb{E}} -\log(\sigma(f(\mathbf{z}_{u},\mathbf{z}_{i})-  
 \alpha_t \cdot w_{u,n} \cdot f(\mathbf{z}_{u},\mathbf{z}_{n}))),
\end{aligned}
\end{equation}
where $\alpha_t=\frac{t}{T}$. The weight factor $w_{u,n}$ is defined as $w_{u,n} = 1 - c_{u,n}$, where $c_{u,n}$ represents the normalized confidence score, calculated by scaling all confidence scores stored in the confidence memory to the range [0,1]. Notably, a higher value of $c_{u,n}$ indicates that the corresponding sample possesses a lower degree of noise.

By incorporating LossEraser with enhancements such as progressive reversal and adaptive reversal, we achieve the goal of eliminating the need for re-training. The final recommendation loss is formulated as:
\begin{equation}
	\label{equation11}
	\begin{split}
		\mathcal{L} = \mathcal{L}_{rec} + \alpha \mathcal{L}_{eraser},
	\end{split}
\end{equation}
where $\alpha$ is the weight control the influence of $\mathcal{L}_{eraser}$.

After completing the LossEraser training process, the agent obtains an updated recommendation model $\theta_{f}^{*}$ and the corresponding new recommendation losses $\mathcal{L}_{rec}$ for each interaction, which serve as the basis for subsequent reasoning. It is also important to note that, even with the inclusion of $\mathcal{L}_{eraser}$, we still record $\mathcal{L}_{rec}$ as the observation loss metric for the agent to determine whether a sample is considered noisy.

\subsubsection{Model Evaluation.}
After multiple LossEraser training cycles, the agent may focus on changes in the performance of the recommendation model. These performance shifts serve as critical metrics, reflecting how well the model adapts to the evolving denoising rules and adjustments. By tracking these shifts, the agent can identify patterns in model behavior and decide whether to update confidence scores, refine denoising rules, or proceed with further model training in the next planning step. To facilitate this, the agent invokes the evaluator to assess the current model performance. The evaluation process can be formulated as follows:
\begin{equation} \begin{aligned} \label{equation9} M^{A} \leftarrow \operatorname{Evaluator}(\theta_{f}^{*}), \end{aligned} \end{equation}
where $\operatorname{Evaluator}$ represents the evaluation of the model with parameters $\theta_{f}^{*}$ on the validation set. The evaluation outcome is subsequently appended as the additional textual description associated with the current action.

After completing each action, RuleAgent employs the ${Planning}$ module again to determine its subsequent action. The agent will automatically terminate and signal the conclusion of the process either after executing a predetermined number of actions or if the model evaluation exhibits a decline over five consecutive iterations. 

\subsection{Output}  Apart from initializing the memory modules, we do not intervene in any part of the process—everything is handled autonomously by the agent through reflection. Compared to traditional methods, where data experts repeatedly evaluate and retrain models to explore denoising rules, our RuleAgent framework is more efficient and convenient. As a result, we can obtain three valuable outcomes: 
\begin{itemize}[leftmargin=12pt]
    \item Hierarchical denoising rules extracted from the rule memory, which provide useful insights for human experts;
    \item An optimal recommendation model trained on the clean interaction data generated by the agent, ready for deployment in recommendation tasks;
    \item Confidence scores with explanations for each sample, which assists researchers in data cleaning tasks. 
\end{itemize}
Additionally, the detailed system prompts and corresponding outputs of the ${Planning}$ process and two reflection processes, ${Reflection}^{C}$ and ${Reflection}^{R}$, can be seen in the Appendix \ref{systemprompt}.

\section{Experiments}
This section tests the effectiveness of RuleAgent. Specifically, we aim to answer the following questions.(Q1): How does the performance of RuleAgent compare to state-of-the-art denoising methods? (Q2): Are the rules discovered by RuleAgent generalizable? (Q3): What impact does the proposed LossEraser have on RuleAgent? (Q4): How do different LLMs impact the performance of RuleAgent? (Q5): How does RuleAgent discover the rule?

\begin{table}[h]
\caption{Statistics of datasets.}\vspace{-1em}
\footnotesize
\label{datasets}
\centering
\begin{tabular}{@{}ccccc@{}}
\toprule
\textbf{Datasets}           & \textbf{\#Users} & \textbf{\#Items} & \textbf{\#Interactions} & \textbf{Density} \\ \midrule
\textbf{Beauty (full)}         & 22,363           & 12,099           & 198,503                 & 0.073\%           \\
-- \textbf{small}                   & 100           & 2,776           & 8,604                 & 3.099\%           \\
\midrule
\textbf{Yelp2018 (full)}      & 31,668           & 38,048           & 1,561,406               & 0.130\%               \\
-- \textbf{small}                   & 100           & 2,882           & 4,662                 & 1.617\%            \\
\midrule
\textbf{Gowalla (full)}      & 18,737          & 32,495           & 741,906               & 0.1218\%           \\
-- \textbf{small}                    & 100           & 2,498           & 4,096                 & 1.639\%           \\
\bottomrule
\end{tabular}
\vspace{-1em}
\end{table}

\begin{table*}[h]
\caption{Comparison of different denoising methods on recommendation tasks, using Recall (R) and NDCG (N) as the metrics. Best results are in bold, and runner-ups are underlined.}
\label{performance comparison}
\vspace{-1em}
\centering
\resizebox{0.95\textwidth}{!}{
\begin{tabular}{@{}c|c|llll|llll|llll@{}}
\toprule
\multicolumn{2}{c|}{Dataset} & \multicolumn{4}{c|}{Beauty-small} & \multicolumn{4}{c|}{Yelp2018-small} & \multicolumn{4}{c}{Gowalla-small} \\ \midrule
\multicolumn{1}{c|}{Base Model} & Method & \multicolumn{1}{c|}{R@10} & \multicolumn{1}{c|}{R@20} & \multicolumn{1}{c|}{N@10} & N@20 & \multicolumn{1}{c|}{R@10} & \multicolumn{1}{c|}{R@20} & \multicolumn{1}{c|}{N@10} & N@20 & \multicolumn{1}{c|}{R@10} & \multicolumn{1}{c|}{R@20} & \multicolumn{1}{c|}{N@10} & N@20 \\ \midrule
\multirow{6}{*}{GMF} 
& Normal & \multicolumn{1}{c|}{0.05960} & \multicolumn{1}{c|}{0.12109} & \multicolumn{1}{c|}{0.14301} & 0.15338 & \multicolumn{1}{c|}{0.00849} & \multicolumn{1}{c|}{0.01666} & \multicolumn{1}{c|}{0.00956} & 0.01265 & \multicolumn{1}{c|}{0.04330} & \multicolumn{1}{c|}{0.04472} & \multicolumn{1}{c|}{0.03332} & 0.03411 \\
& T-CE & \multicolumn{1}{c|}{\underline{0.07074}} & \multicolumn{1}{c|}{0.12603} & \multicolumn{1}{c|}{\underline{0.16938}} & 0.15841 & \multicolumn{1}{c|}{\underline{0.01066}} & \multicolumn{1}{c|}{\underline{0.03305}} & \multicolumn{1}{c|}{0.01247} & 0.01867 & \multicolumn{1}{c|}{0.04823} & \multicolumn{1}{c|}{0.04831} & \multicolumn{1}{c|}{0.03557} & 0.03526 \\
& DeCA & \multicolumn{1}{c|}{0.06981} & \multicolumn{1}{c|}{0.13782} & \multicolumn{1}{c|}{0.16836} & 0.15987 & \multicolumn{1}{c|}{0.00965} & \multicolumn{1}{c|}{0.02861} & \multicolumn{1}{c|}{\underline{0.01686}} & \underline{0.02364} & \multicolumn{1}{c|}{\underline{0.05133}} & \multicolumn{1}{c|}{0.04862} & \multicolumn{1}{c|}{\underline{0.03978}} & \underline{0.03737} \\
& BOD & \multicolumn{1}{c|}{0.06812} & \multicolumn{1}{c|}{\underline{0.13906}} & \multicolumn{1}{c|}{0.16758} & 0.16665 & \multicolumn{1}{c|}{0.00858} & \multicolumn{1}{c|}{0.02152} & \multicolumn{1}{c|}{0.01595} & 0.01462 & \multicolumn{1}{c|}{0.04597} & \multicolumn{1}{c|}{0.04692} & \multicolumn{1}{c|}{0.03837} & 0.03614 \\
& DCF & \multicolumn{1}{c|}{0.06919} & \multicolumn{1}{c|}{0.13496} & \multicolumn{1}{c|}{0.16695} & \underline{0.16707} & \multicolumn{1}{c|}{0.00946} & \multicolumn{1}{c|}{0.01982} & \multicolumn{1}{c|}{0.01327} & 0.01683 & \multicolumn{1}{c|}{0.04387} & \multicolumn{1}{c|}{\underline{0.04952}} & \multicolumn{1}{c|}{0.03347} & 0.03533 \\ \cmidrule(l){2-14}
& RuleAgent & \multicolumn{1}{c|}{\textbf{0.07744}} & \multicolumn{1}{c|}{\textbf{0.14296}} & \multicolumn{1}{c|}{\textbf{0.17246}} & \textbf{0.16779} & \multicolumn{1}{c|}{\textbf{0.02808}} & \multicolumn{1}{c|}{\textbf{0.04643}} & \multicolumn{1}{c|}{\textbf{0.02683}} & \textbf{0.03335} & \multicolumn{1}{c|}{\textbf{0.05583}} & \multicolumn{1}{c|}{\textbf{0.08912}} & \multicolumn{1}{c|}{\textbf{0.04505}} & \textbf{0.05364} \\ \midrule
\multirow{6}{*}{LightGCN} 
& Normal & \multicolumn{1}{c|}{0.07753} & \multicolumn{1}{c|}{0.14056} & \multicolumn{1}{c|}{0.17422} & 0.16844 & \multicolumn{1}{c|}{0.04307} & \multicolumn{1}{c|}{0.05524} & \multicolumn{1}{c|}{0.03257} & 0.03851 & \multicolumn{1}{c|}{0.05898} & \multicolumn{1}{c|}{0.09652} & \multicolumn{1}{c|}{0.04501} & 0.05872 \\
& T-CE & \multicolumn{1}{c|}{0.07815} & \multicolumn{1}{c|}{0.14333} & \multicolumn{1}{c|}{0.17726} & 0.17022 & \multicolumn{1}{c|}{0.04432} & \multicolumn{1}{c|}{0.05729} & \multicolumn{1}{c|}{0.03472} & 0.04022 & \multicolumn{1}{c|}{0.06244} & \multicolumn{1}{c|}{\underline{0.10608}} & \multicolumn{1}{c|}{0.04639} & 0.05955 \\
& DeCA & \multicolumn{1}{c|}{\underline{0.07915}} & \multicolumn{1}{c|}{\underline{0.14632}} & \multicolumn{1}{c|}{0.18323} & 0.17321 & \multicolumn{1}{c|}{0.04607} & \multicolumn{1}{c|}{0.05879} & \multicolumn{1}{c|}{\underline{0.04225}} & 0.04782 & \multicolumn{1}{c|}{\underline{0.07374}} & \multicolumn{1}{c|}{0.09808} & \multicolumn{1}{c|}{\underline{0.05984}} & 0.05903 \\
& BOD & \multicolumn{1}{c|}{0.07832} & \multicolumn{1}{c|}{0.14402} & \multicolumn{1}{c|}{\underline{0.18732}} & \underline{0.17812} & \multicolumn{1}{c|}{\underline{0.05041}} & \multicolumn{1}{c|}{0.05681} & \multicolumn{1}{c|}{0.03725} & \underline{0.04802} & \multicolumn{1}{c|}{0.06192} & \multicolumn{1}{c|}{0.09913} & \multicolumn{1}{c|}{0.05366} & 0.06132 \\
& DCF & \multicolumn{1}{c|}{0.07845} & \multicolumn{1}{c|}{0.13902} & \multicolumn{1}{c|}{0.17464} & 0.16705 & \multicolumn{1}{c|}{0.04507} & \multicolumn{1}{c|}{\underline{0.05952}} & \multicolumn{1}{c|}{0.03601} & 0.04171 & \multicolumn{1}{c|}{0.06754} & \multicolumn{1}{c|}{0.10152} & \multicolumn{1}{c|}{0.05287} & \underline{0.06502} \\ \cmidrule(l){2-14}
& RuleAgent & \multicolumn{1}{c|}{\textbf{0.08368}} & \multicolumn{1}{c|}{\textbf{0.15113}} & \multicolumn{1}{c|}{\textbf{0.18914}} & \textbf{0.18264} & \multicolumn{1}{c|}{\textbf{0.06888}} & \multicolumn{1}{c|}{\textbf{0.08782}} & \multicolumn{1}{c|}{\textbf{0.04897}} & \textbf{0.05958} & \multicolumn{1}{c|}{\textbf{0.08634}} & \multicolumn{1}{c|}{\textbf{0.10938}} & \multicolumn{1}{c|}{\textbf{0.06092}} & \textbf{0.06921} \\ \midrule
\multirow{6}{*}{XSimGCL} 
& Normal & \multicolumn{1}{c|}{0.08101} & \multicolumn{1}{c|}{0.15184} & \multicolumn{1}{c|}{0.19458} & 0.18733 & \multicolumn{1}{c|}{0.05075} & \multicolumn{1}{c|}{0.06464} & \multicolumn{1}{c|}{0.04171} & 0.04653 & \multicolumn{1}{c|}{0.06193} & \multicolumn{1}{c|}{0.10155} & \multicolumn{1}{c|}{0.04871} & 0.06234 \\
& T-CE & \multicolumn{1}{c|}{0.08192} & \multicolumn{1}{c|}{0.15227} & \multicolumn{1}{c|}{\underline{0.19967}} & 0.18862 & \multicolumn{1}{c|}{\underline{0.05847}} & \multicolumn{1}{c|}{0.06843} & \multicolumn{1}{c|}{0.04342} & 0.04884 & \multicolumn{1}{c|}{0.06887} & \multicolumn{1}{c|}{{0.10711}} & \multicolumn{1}{c|}{0.05303} & 0.06610 \\
& DeCA & \multicolumn{1}{c|}{\underline{0.08332}} & \multicolumn{1}{c|}{{0.15341}} & \multicolumn{1}{c|}{0.19816} & 0.18815 & \multicolumn{1}{c|}{{0.05788}} & \multicolumn{1}{c|}{0.06544} & \multicolumn{1}{c|}{\underline{0.04881}} & 0.04910 & \multicolumn{1}{c|}{{0.07398}} & \multicolumn{1}{c|}{0.10645} & \multicolumn{1}{c|}{\underline{0.06250}} & 0.06421 \\
& BOD & \multicolumn{1}{c|}{0.08273} & \multicolumn{1}{c|}{\underline{0.15508}} & \multicolumn{1}{c|}{{0.19693}} & \underline{0.18955} & \multicolumn{1}{c|}{{0.05767}} & \multicolumn{1}{c|}{\underline{0.06855}} & \multicolumn{1}{c|}{0.04715} & \underline{0.04939} & \multicolumn{1}{c|}{\underline{0.07767}} & \multicolumn{1}{c|}{0.10368} & \multicolumn{1}{c|}{0.05603} & \underline{0.06983} \\
& DCF & \multicolumn{1}{c|}{0.08214} & \multicolumn{1}{c|}{0.15461} & \multicolumn{1}{c|}{0.19511} & 0.18823 & \multicolumn{1}{c|}{0.05375} & \multicolumn{1}{c|}{{0.06739}} & \multicolumn{1}{c|}{0.04688} & 0.05091 & \multicolumn{1}{c|}{{0.06902}} & \multicolumn{1}{c|}{\underline{0.10908}} & \multicolumn{1}{c|}{0.05642} & {0.06770} \\ \cmidrule(l){2-14}
& RuleAgent & \multicolumn{1}{c|}{\textbf{0.08502}} & \multicolumn{1}{c|}{\textbf{0.15635}} & \multicolumn{1}{c|}{\textbf{0.20832}} & \textbf{0.19419} & \multicolumn{1}{c|}{\textbf{0.06296}} & \multicolumn{1}{c|}{\textbf{0.06951}} & \multicolumn{1}{c|}{\textbf{0.05361}} & \textbf{0.05692} & \multicolumn{1}{c|}{\textbf{0.08798}} & \multicolumn{1}{c|}{\textbf{0.11645}} & \multicolumn{1}{c|}{\textbf{0.06502}} & \textbf{0.07421} \\ 
 \bottomrule
\end{tabular}}
\label{tab:performance_comparison}
\end{table*}

\begin{table*}[h]
\caption{Comparison of RuleAgent-extracted rules and existing denoising methods on recommendation tasks, using Recall (R) and NDCG (N) as the metrics. Best results are in bold, and runner-ups are underlined. }
\label{rule method}
\vspace{-1em}
\centering
\resizebox{0.95\textwidth}{!}{
\begin{tabular}{@{}cc|cccc|cccc|cccc@{}}
\toprule
\multicolumn{2}{c|}{Dataset} &
  \multicolumn{4}{c|}{Beauty} &
  \multicolumn{4}{c|}{Yelp2018} &
  \multicolumn{4}{c}{Gowalla} \\ \midrule
\multicolumn{1}{c|}{Base Model} &
  Method &
  \multicolumn{1}{c|}{R@10} &
  \multicolumn{1}{c|}{R@20} &
  \multicolumn{1}{c|}{N@10} &
  N@20 &
  \multicolumn{1}{c|}{R@10} &
  \multicolumn{1}{c|}{R@20} &
  \multicolumn{1}{c|}{N@10} &
  N@20 &
  \multicolumn{1}{c|}{R@10} &
  \multicolumn{1}{c|}{R@20} &
  \multicolumn{1}{c|}{N@10} &
  N@20 \\ \midrule
\multicolumn{1}{c|}{\multirow{6}{*}{GMF}} &
  Normal &
  \multicolumn{1}{c|}{0.06915} &
  \multicolumn{1}{c|}{0.09968} &
  \multicolumn{1}{c|}{0.04593} &
  0.05554 &
  \multicolumn{1}{c|}{0.02421} &
  \multicolumn{1}{c|}{0.04272} &
  \multicolumn{1}{c|}{0.02667} &
  0.03379 &
  \multicolumn{1}{c|}{0.05303} &
  \multicolumn{1}{c|}{0.08473} &
  \multicolumn{1}{c|}{0.05456} &
  0.06660 \\
\multicolumn{1}{c|}{} &
  T-CE &
  \multicolumn{1}{c|}{0.07094} &
  \multicolumn{1}{c|}{{\ul 0.10163}} &
  \multicolumn{1}{c|}{0.04749} &
  0.05719 &
  \multicolumn{1}{c|}{0.02454} &
  \multicolumn{1}{c|}{0.04322} &
  \multicolumn{1}{c|}{0.02725} &
  0.03392 &
  \multicolumn{1}{c|}{0.05319} &
  \multicolumn{1}{c|}{0.08507} &
  \multicolumn{1}{c|}{0.05471} &
  0.06746 \\
\multicolumn{1}{c|}{} &
  DeCA &
  \multicolumn{1}{c|}{0.07087} &
  \multicolumn{1}{c|}{0.10016} &
  \multicolumn{1}{c|}{0.04732} &
  0.05775 &
  \multicolumn{1}{c|}{0.02482} &
  \multicolumn{1}{c|}{0.04439} &
  \multicolumn{1}{c|}{{\ul 0.02813}} &
  0.03488 &
  \multicolumn{1}{c|}{0.05564} &
  \multicolumn{1}{c|}{0.08722} &
  \multicolumn{1}{c|}{{\ul 0.06193}} &
  {\ul 0.07432} \\
\multicolumn{1}{c|}{} &
  BOD &
  \multicolumn{1}{c|}{{\ul 0.07146}} &
  \multicolumn{1}{c|}{0.09980} &
  \multicolumn{1}{c|}{{\ul 0.04759}} &
  {\ul 0.05812} &
  \multicolumn{1}{c|}{\textbf{0.02513}} &
  \multicolumn{1}{c|}{\textbf{0.04501}} &
  \multicolumn{1}{c|}{0.02802} &
  {\ul 0.03502} &
  \multicolumn{1}{c|}{0.05424} &
  \multicolumn{1}{c|}{0.08532} &
  \multicolumn{1}{c|}{0.05958} &
  0.07382 \\
\multicolumn{1}{c|}{} &
  DCF &
  \multicolumn{1}{c|}{0.06936} &
  \multicolumn{1}{c|}{0.09989} &
  \multicolumn{1}{c|}{0.04692} &
  0.05649 &
  \multicolumn{1}{c|}{0.02462} &
  \multicolumn{1}{c|}{0.04319} &
  \multicolumn{1}{c|}{0.02783} &
  0.03479 &
  \multicolumn{1}{c|}{{\ul 0.05643}} &
  \multicolumn{1}{c|}{\textbf{0.09184}} &
  \multicolumn{1}{c|}{0.05888} &
  0.07182 \\ \cmidrule(l){2-14} 
\multicolumn{1}{c|}{} &
  RuleAgent &
  \multicolumn{1}{c|}{\textbf{0.07351}} &
  \multicolumn{1}{c|}{\textbf{0.10579}} &
  \multicolumn{1}{c|}{\textbf{0.0484}} &
  \textbf{0.05845} &
  \multicolumn{1}{c|}{{\ul 0.02497}} &
  \multicolumn{1}{c|}{{\ul 0.04488}} &
  \multicolumn{1}{c|}{\textbf{0.02976}} &
  \textbf{0.03678} &
  \multicolumn{1}{c|}{\textbf{0.05828}} &
  \multicolumn{1}{c|}{{\ul 0.09045}} &
  \multicolumn{1}{c|}{\textbf{0.06562}} &
  \textbf{0.07688} \\ \midrule
\multicolumn{1}{c|}{\multirow{6}{*}{LightGCN}} &
  Normal &
  \multicolumn{1}{c|}{0.09203} &
  \multicolumn{1}{c|}{0.12982} &
  \multicolumn{1}{c|}{0.06255} &
  0.07435 &
  \multicolumn{1}{c|}{0.03414} &
  \multicolumn{1}{c|}{0.05842} &
  \multicolumn{1}{c|}{0.03934} &
  0.04809 &
  \multicolumn{1}{c|}{0.07096} &
  \multicolumn{1}{c|}{0.10892} &
  \multicolumn{1}{c|}{0.07809} &
  0.09141 \\
\multicolumn{1}{c|}{} &
  T-CE &
  \multicolumn{1}{c|}{0.09373} &
  \multicolumn{1}{c|}{0.13268} &
  \multicolumn{1}{c|}{0.06298} &
  0.07515 &
  \multicolumn{1}{c|}{{\ul 0.03478}} &
  \multicolumn{1}{c|}{0.05882} &
  \multicolumn{1}{c|}{0.03952} &
  {\ul 0.04892} &
  \multicolumn{1}{c|}{0.07121} &
  \multicolumn{1}{c|}{0.10895} &
  \multicolumn{1}{c|}{0.07892} &
  0.09206 \\
\multicolumn{1}{c|}{} &
  DeCA &
  \multicolumn{1}{c|}{0.09345} &
  \multicolumn{1}{c|}{0.13211} &
  \multicolumn{1}{c|}{0.06248} &
  0.07534 &
  \multicolumn{1}{c|}{0.03444} &
  \multicolumn{1}{c|}{0.05872} &
  \multicolumn{1}{c|}{0.03962} &
  0.04874 &
  \multicolumn{1}{c|}{0.07142} &
  \multicolumn{1}{c|}{0.10942} &
  \multicolumn{1}{c|}{0.07888} &
  0.09196 \\
\multicolumn{1}{c|}{} &
  BOD &
  \multicolumn{1}{c|}{{\ul 0.09388}} &
  \multicolumn{1}{c|}{0.13341} &
  \multicolumn{1}{c|}{\underline{0.06313}} &
  \underline{0.07584} &
  \multicolumn{1}{c|}{0.03462} &
  \multicolumn{1}{c|}{0.05893} &
  \multicolumn{1}{c|}{0.03958} &
  0.04821 &
  \multicolumn{1}{c|}{0.07168} &
  \multicolumn{1}{c|}{0.10941} &
  \multicolumn{1}{c|}{0.07884} &
  0.09213 \\
\multicolumn{1}{c|}{} &
  DCF &
  \multicolumn{1}{c|}{0.09336} &
  \multicolumn{1}{c|}{{\ul 0.13353}} &
  \multicolumn{1}{c|}{0.06262} &
  0.07511 &
  \multicolumn{1}{c|}{0.03454} &
  \multicolumn{1}{c|}{{\ul 0.05913}} &
  \multicolumn{1}{c|}{{\ul 0.03964}} &
  0.04862 &
  \multicolumn{1}{c|}{{\ul 0.07178}} &
  \multicolumn{1}{c|}{{\ul 0.11043}} &
  \multicolumn{1}{c|}{{\ul 0.07904}} &
  {\ul 0.09249} \\ \cmidrule(l){2-14} 
\multicolumn{1}{c|}{} &
  RuleAgent &
  \multicolumn{1}{c|}{\textbf{0.09416}} &
  \multicolumn{1}{c|}{\textbf{0.13358}} &
  \multicolumn{1}{c|}{\textbf{0.06378}} &
  \textbf{0.07651} &
  \multicolumn{1}{c|}{\textbf{0.03510}} &
  \multicolumn{1}{c|}{\textbf{0.06012}} &
  \multicolumn{1}{c|}{\textbf{0.04012}} &
  \textbf{0.04918} &
  \multicolumn{1}{c|}{\textbf{0.07223}} &
  \multicolumn{1}{c|}{\textbf{0.11342}} &
  \multicolumn{1}{c|}{\textbf{0.08052}} &
  \textbf{0.09288} \\ \bottomrule
\end{tabular}}
\label{tab:performance_comparison}
\end{table*}

\subsection{Experimental Settings}
\textbf{Datasets.} We use three commonly used public benchmark datasets in our experiments: Beauty\footnote{\url{https://jmcauley.ucsd.edu/data/amazon/links.html}}
, Gowalla\footnote{\url{http://snap.stanford.edu/data/loc-gowalla.html}}, and Yelp2018\footnote{\url{https://www.yelp.com/dataset}}. Due to the high cost of API calls, we follow existing studies~\cite{68wang2023user,71zhang2024agentcf} to sample dense subsets from these datasets. Each subset includes 100 users with the most interactions, offering more data for analysis. The dataset statistics are presented in Table ~\ref{datasets}.

\stitle{Evaluation Metrics.} We divide the datasets into three parts: training set, validation set, and test set, following a 7:1:2 ratio. Two commonly used evaluation metrics, Recall$@K$ and NDCG$@K$, are employed with $K$ set to 10 and 20. Each metric is calculated 10 times, and the average results are reported.

\stitle{Baselines.} The primary goal of this paper is to denoise feedback to enhance the performance of existing base recommendation models. To achieve this, we select three commonly used implicit feedback-based recommendation models as the base models for denoising:
\begin{itemize}[leftmargin=12pt]
    \item GMF~\cite{39he2017neural}: A generalized matrix factorization-based model.
    \item LightGCN~\cite{26he2020lightgcn}: A graph-based model that simplifies GCN design for recommendation by removing nonlinear  transformations.
    \item XSimGCL \cite{82yu2023xsimgcl}: A fast contrastive learning recommendation method that streamlines contrastive view construction and optimizes representation uniformity.
\end{itemize}

\vspace{0.5em}
To evaluate the denoising effect, we select five existing denoising methods to compare their performance with the aforementioned recommendation models:
\begin{itemize}[leftmargin=12pt]
    \item T-CE (2021)~\cite{11wang2021denoising}: Removes samples with loss values exceeding a predefined threshold.
    \item DeCA (2022)~\cite{10wang2022learning}: Combines predictions from two different models to identify noisy samples based on their disagreement.
    \item BOD (2023)~\cite{65wang2023efficient}: Automatically learns sample weights through bi-level optimization.
    \item DCF (2024)~\cite{66he2024double}: Introduces a double correction mechanism to iteratively relabel samples, addressing both noise issues and data sparsity simultaneously.
\end{itemize}

\stitle{Parameter Settings.} We implement RuleAgent using PyTorch, and the reflection and inference of RuleAgent are implemented using GPT-4o mini\footnote{\url{https://openai.com/index/gpt-4o-mini-advancing-cost-efficient-intelligence/}}. For all methods, we follow the recommended parameter settings unless stated otherwise. Specifically, we set the batch size to 512, the learning rate to 0.001, the embedding size to 64, and the number of LightGCN and XSimGCL layers to 2. The models are optimized using the Adam optimizer~\cite{41kingma2014adam}, and the Xavier initializer~\cite{42glorot2010understanding} is used for parameter initialization.

\subsection{Performance Comparison (Q1)}
We first compare RuleAgent with existing denoising methods on three small datasets. The results, shown in Table \ref{performance comparison}, allow us to draw the following conclusions. The proposed RuleAgent significantly enhances the performance of the three base models, achieving the best or second-best results in all cases. We attribute these improvements to RuleAgent's ability to continuously update denoising rules, tailoring them to each dataset. Unlike existing denoising methods, which rely on fixed rules that constrain their performance to specific datasets, RuleAgent extracts dataset-specific denoising rules, providing a notable advantage. Furthermore, we observe that the second-best method varies across different datasets and base models, underscoring the limitations of fixed rules in adapting to diverse scenarios. Additionally, all denoising approaches outperform normal training, reinforcing the importance of denoising in recommender systems. These findings align with prior studies~\cite{10wang2022learning,12DBLP:conf/sigir/Gao0HCZFZ22}.

\begin{figure}[t]
  \centering
  \includegraphics[width=1\linewidth]{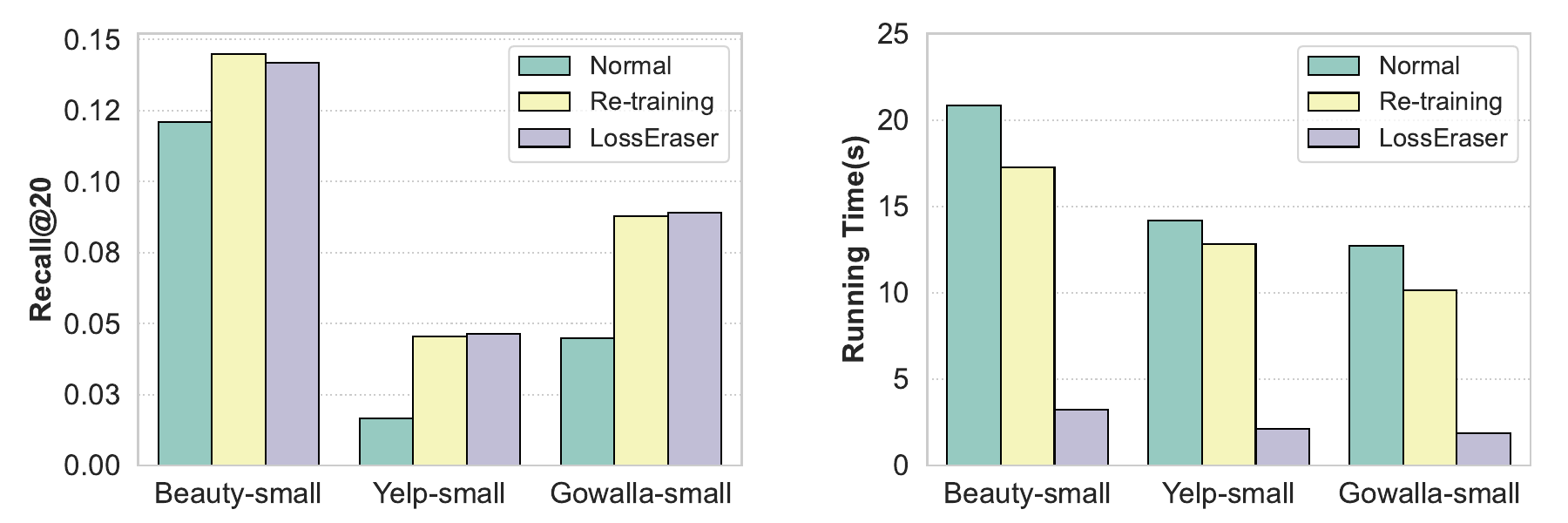}
  \caption{The Comparison of Different Training Strategies.}
  \vspace{-1em}
\label{lossearser}
\end{figure}

\subsection{Effectiveness of Discovered Rules (Q2)}
To validate the generalizability of RuleAgent, we further evaluate the effectiveness of the denoising rules extracted by the framework. Specifically, we utilize ChatGPT-4o-with-Canvas\footnote{\url{https://openai.com/index/chatgpt/}} to summarize the contents of the denoising rule memory and translate the summarized rules into implementation code for the recommendation model. We apply these new denoising rules to various open-source datasets and compare their performance with the latest denoising methods. Specifically, we use the discovered rules to filter the samples, and during the loss calculation, we zero out the loss corresponding to the filtered samples, while other denoising methods process the samples using the techniques described in their respective papers. From the results in Table \ref{rule method}, we observe that the denoising methods derived from the rules discovered by RuleAgent consistently achieved the best or second-best performance in all cases, which demonstrates that the denoising rules discovered by RuleAgent are effective. Therefore, although the cost remains a current limitation of RuleAgent, a feasible approach could be to allow RuleAgent to reason over a small set of domain-specific data and extract effective denoising rules that can then be applied to the entire dataset. It is worth noting that the results of XSimGCL across all datasets also align with the above conclusions; however, due to space limitations, they are not presented in Table \ref{rule method}.

\vspace{-1em}
\subsection{Analysis of LossEraser (Q3)}
We use GMF as the base model and report the recommendation performance and real running time for different training strategies over a complete training cycle. We set the parameter \(\alpha\) to 0.01 (due to space constraints, we do not present the sensitivity analysis of \(\alpha\), but based on the experimental results, \(\alpha\) should lie between 0.01 and 0.1; values outside this range lead to obviously degraded performance). The results shown in Figure~\ref{lossearser} were collected on an Intel(R) Core(TM) i9-10900X CPU and a GeForce RTX 3090Ti GPU. As seen in Figure~\ref{lossearser}, RuleAgent outperforms normal training in terms of recommendation performance, whether using re-training or the LossEraser strategy. Since RuleAgent drops some data, the re-training dataset is smaller than that of normal training, resulting in reduced training time. Furthermore, LossEraser reduces the training time by approximately four times compared to re-training, while achieving nearly the same recommendation performance, demonstrating the effectiveness of the LossEraser strategy.

\subsection{Comparison of Different Backbones (Q4)}
In addition to GPT-4o mini, we evaluate the performance of other LLMs as the core reasoning backbones for RuleAgent, including Gemma-2B\footnote{\url{https://huggingface.co/google/gemma-1.1-2b-it}}, Llama3-8B\footnote{\url{https://huggingface.co/meta-llama/Meta-Llama-3-8B}}, Qwen-Turbo\footnote{\url{https://tongyi.aliyun.com/qianwen/}}, and GLM-4-Flash\footnote{https://open.bigmodel.cn/dev/api/normal-model/glm-4}. The experimental results in Table \ref{llm}, reveal that RuleAgent achieves the strongest performance when powered by GPT-4o mini, although its advantage over Qwen-Turbo and GLM-4-Flash is marginal. Conversely, smaller LLMs, such as Gemma-2B, exhibit a substantial performance degradation. This discrepancy highlights the critical role of reasoning capabilities in our task, where larger LLMs demonstrate a pronounced advantage over their smaller counterparts. Based on these findings, we recommend employing large-scale models as the backbone to ensure optimal performance.

\begin{table}[h]
\caption{Performance comparison of different backbones.}
\label{llm}
\vspace{-1em}
\centering
\resizebox{0.48\textwidth}{!}{
\begin{tabular}{@{}c|ll|ll|ll@{}}
\toprule
\multirow{2}{*}{LLM} & \multicolumn{2}{c|}{Beauty-small}      & \multicolumn{2}{c|}{Yelp-small}        & \multicolumn{2}{c}{Gowalla-small}      \\ \cmidrule(l){2-7} 
 &
  \multicolumn{1}{c|}{R@20} &
  \multicolumn{1}{c|}{N@20} &
  \multicolumn{1}{c|}{R@20} &
  \multicolumn{1}{c|}{N@20} &
  \multicolumn{1}{c|}{R@20} &
  \multicolumn{1}{c}{N@20} \\ \midrule
Gemma-2B             & \multicolumn{1}{l|}{0.12551} & 0.15481 & \multicolumn{1}{l|}{0.02136} & 0.01466 & \multicolumn{1}{l|}{0.04615} & 0.03711 \\
Llama3-8B            & \multicolumn{1}{l|}{0.13350} & 0.16015 & \multicolumn{1}{l|}{0.03215} & 0.01948 & \multicolumn{1}{l|}{0.05218} & 0.04815 \\
Qwen-Turbo        & \multicolumn{1}{l|}{0.14084} & 0.16636 & \multicolumn{1}{l|}{0.04582} & 0.03126 & \multicolumn{1}{l|}{0.08248} & 0.05215 \\
GLM-4-Flash          & \multicolumn{1}{l|}{0.13945} & 0.16451 & \multicolumn{1}{l|}{0.04445} & 0.03248 & \multicolumn{1}{l|}{0.08465} & 0.05162 \\
GPT-4o mini &
  \multicolumn{1}{l|}{\textbf{0.14196}} &
  \textbf{0.16779} &
  \multicolumn{1}{l|}{\textbf{0.04643}} &
  \textbf{0.03335} &
  \multicolumn{1}{l|}{\textbf{0.08910}} &
  \textbf{0.05362} \\ \bottomrule
\end{tabular}}
\vspace{-1em}
\end{table}



\subsection{Analysis of Rule Discovery Process (Q5)}
\begin{figure}[h]
\setlength{\belowcaptionskip}{-15pt}
  \centering
  \includegraphics[width=0.98\linewidth]{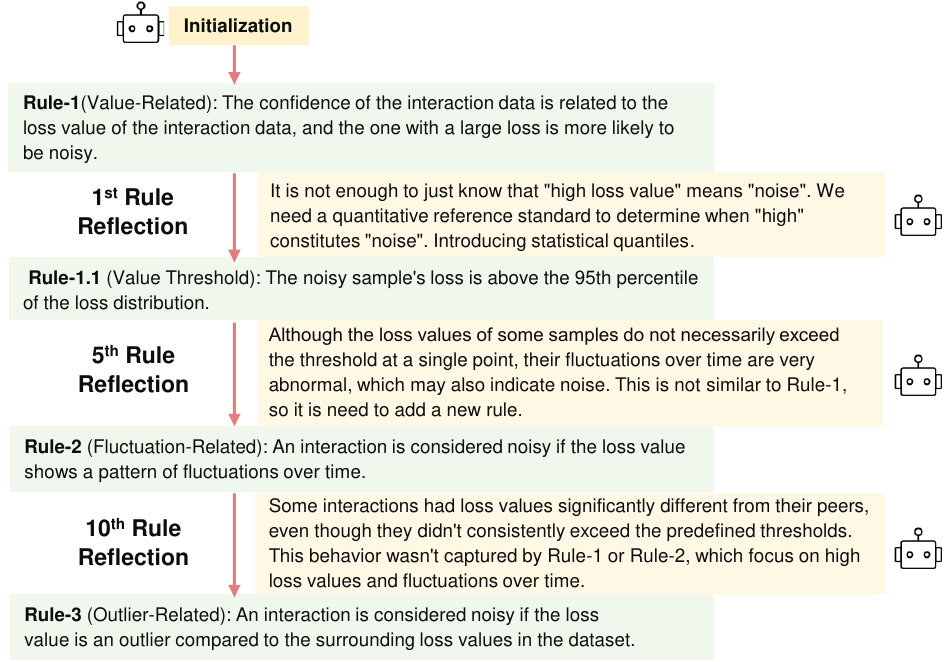}
  \caption{Rule Discovery Process.}
\label{rule discovery}
\end{figure}

We record the process by which RuleAgent discovers denoising rules. Figure \ref{rule discovery} illustrates part of this process using the base model GMF on the Beauty dataset. In the first round of rule reflection, RuleAgent autonomously determines the need for a 95\% threshold in the initial Rule-1, revealing a human-like line of reasoning. After multiple iterations, during the fifth rule reflection, RuleAgent recognizes that the existing rules are inadequate for characterizing noisy samples and thus introduces a new rule. Due to space constraints, we do not present all rule-reflection outcomes; however, in additional rounds, RuleAgent refines or merges existing rules, demonstrating powerful autonomous reasoning ability.

\section{Related Work}

\subsection{Recommendation Denoising} 
A common approach to denoising implicit feedback is to identify clean and informative samples and use them to train a recommendation model. Following this idea, WBPR~\cite{08gantner2012personalized} treats popular items without interactions as true negatives and samples them more frequently. Recent studies have revealed that the noise level of samples often correlates with training loss. T-CE~\cite{11wang2021denoising}, one of the earliest works in this area, identifies samples with high loss as more likely to be noisy. Similarly, DeCA~\cite{10wang2022learning} suggests that clean samples produce similar predictions across different models. By jointly training two recommendation models, it differentiates clean samples from noisy ones based on prediction disagreements. SGDL~\cite{12DBLP:conf/sigir/Gao0HCZFZ22} argues that clean samples are often identified during the initial training stages and proposes adaptively weighting samples based on their similarity. BOD~\cite{65wang2023efficient} further posits that samples yielding consistent losses across different loss functions are likely to be clean. Additionally, DCF~\cite{66he2024double} addresses both denoising and data sparsity issues through a double correction strategy that relabels observed and unobserved samples to achieve denoising.


\subsection{Language Agents in Recommendation} 
LLM-powered agents have recently gained significant attention due to their strong reasoning and planning capabilities, making them promising tools for developing artificial general intelligence (AGI)~\cite{85argyle2023out,86rana2023sayplan}. A key advancement has been the introduction of agents into recommendation systems, aimed at either enhancing recommendation performance or transforming the recommendation paradigm. For example, InteRecAgent~\cite{69huang2023recommender} employs an LLM as its core engine and integrates recommendation models as tools, creating a conversational recommendation system with a natural language interface. Similarly, RecMind~\cite{70wang2023recmind} is an autonomous recommendation agent powered by an LLM that utilizes external knowledge and well-designed tools to provide zero-shot personalized recommendations. AgentCF~\cite{71zhang2024agentcf} proposes an agent-based collaborative filtering approach, where user and item agents model their bilateral relationships, replacing traditional collaborative filtering techniques and introducing a novel recommendation paradigm.

Some studies have gone beyond improving recommendations by leveraging agents to simulate user behavior. RecAgent~\cite{68wang2023user} and Agent4Rec~\cite{72zhang2024generative} adopt multiple agents as user simulators in recommendation systems. These generative agents, powered by LLM, are specifically designed for recommendation scenarios with tailored components such as user profiles, memory, and operation modules. They excel at reproducing intricate behaviors, including users’ day-to-day interactions, item engagements, and even social dynamics between users in realistic settings.

\section{Conclusion}
In this paper, we explore the potential of autonomous LLM-powered agents to discover rules for recommendation denoising. The proposed RuleAgent is equipped with several modules, effectively simulating the role of a data denoising expert and uncovering rules. Extensive experiments on three real-world datasets demonstrate the superior performance of RuleAgent. 

In light of RuleAgent's demonstrated capabilities in leveraging the agent-based rule discovery paradigm to address data denoising challenges, it holds significant potential as a valuable tool for improving data quality across a wide range of applications, extending beyond recommender systems. In future work, we aim to incorporate multi-agent collaborative reasoning to further enhance its effectiveness, while also broadening its validation to encompass a wider variety of data scenarios.



\balance
\bibliographystyle{ACM-Reference-Format}
\bibliography{RuleAgent}

\clearpage
\appendix

\section{APPENDIX}
\subsection{System Prompt Examples}
\label{systemprompt}
\stitle{System Prompt of $Planning$.}\vspace{3pt}\\
\setlength{\fboxsep}{5pt}
\colorbox{bgcolor}{
  \begin{varwidth}{\dimexpr\linewidth-2\fboxsep\relax}
  \small
  \emph{
 You have three potential planning paths for deciding the next step:
    1. Current Confidence-Based Planning: Current confidence scores are \textcolor{purple}{<$M^{C}$>}. Examine the confidence levels and their associated reasons stored in your confidence data. Compare these confidence levels with the model’s training results. Determine the most suitable next step and explain your reasoning.\\
    2. Current Rule-Based Planning: Current rule \textcolor{purple}{<$M^{R}$>}. Refer to the rules stored in your current rule set. Compare the model training results against these rules. Decide on the most appropriate next step and provide the rationale.\\
    3. Historical Action-Based Planning: The historical actions taken are \textcolor{purple}{<$M^{A}$>}. Analyze past actions and their outcomes. Use these insights to determine the most suitable next step and explain why.\\    
    The available actions are:\\
  a. Confidence Reflection. This action is to update the confidence score and corresponding explanation of samplings. You will choose this action when the current confidence scores assigned to the sampled data are found to be inaccurate or insufficient.\\
  b. Rule Reflection. This action is to update denoising rule. You will choose this action when the existing denoising rule is either incorrect or can be further refined.\\
  c. LossEraser Training. Use the confidence memory to filter data for the next recommendation training phase, then you can get new recommendation model parameters.\\
  d. Model Evaluation. Evaluate current performance of recommendation model to show whether it has improved or worsened.\\
  Compare the decisions with these three planning path, then decide on the most suitable next step and explain your reasoning.\\
  Please provide a response that strictly follows the response format:\\
  The next action is: \color{RoyalBlue}{<a/b/c/d>}. The reason for this decision is: \color{RoyalBlue}{<Your Explanation>}.
  }
  \end{varwidth}}
  
\stitle{Output Example of $Planning$.}\vspace{3pt}\\
\setlength{\fboxsep}{5pt}
\colorbox{bgcolor}{
  \begin{varwidth}{\dimexpr\linewidth-2\fboxsep\relax}
  \small
  \emph{
    The next action is: \uline{a}, the reason for this decision is: \uline{The \textcolor{red}{historical actions} taken so far have focused on initializing the model with the latest user-item interactions (Action c) and evaluating the model's performance post-initialization (Action d). With the model's performance having improved after the initialization, the next step should be to update the \textcolor{red}{confidence scores} (Action a). This action is suitable because it addresses the need to refine and possibly correct the initial confidence scores assigned to the sampled data based on the performance outcomes. It also ensures a solid foundation for further adjustments and refinements in the \textcolor{red}{denoising rule} (Action b), if needed.}
  }
  \end{varwidth}}

\stitle{System Prompt of $Reflection^{C}$.}\vspace{3pt}\\
\setlength{\fboxsep}{5pt}
\colorbox{bgcolor}{
  \begin{varwidth}{\dimexpr\linewidth-2\fboxsep\relax}
    \small
  \emph{
    Please provide the updated confidence score (0-2, 0-1 means noisy sample, 1-2 means clean sample) and explain why you assigned this updated score. 
    The denoising rule is: \textcolor{purple}{<$M^{R}$>}, and User Index is \textcolor{olive}{<UserIndex>}, Item Index is \textcolor{teal}{<ItemIndex>}, Historical Loss is \textcolor{purple}{<$\mathcal{L}$>}. 
    You previously provided: The confidence score is \textcolor{purple}{<$Score^{C}$>}. The explanation was: \textcolor{purple}{<$Reason^{C}$>}.
    Please analyze the reason and update your confidence score and explanation accordingly.\\
    The response format should be: The confidence score is \color{RoyalBlue}{<0-2>}. The explanation: \color{RoyalBlue}{<Your Explanation>}.
    }
  \end{varwidth}}

\stitle{Output Example of $Reflection^{C}$.}\vspace{3pt}\\
\setlength{\fboxsep}{5pt}
\colorbox{bgcolor}{
  \begin{varwidth}{\dimexpr\linewidth-2\fboxsep\relax}
  \small
  \emph{
    The confidence score is \uline{0.125}. The explanation: \uline{The updated confidence score has been significantly reduced from 0.25 to 0.125 due to the persistent and substantial negative impact on the model's performance that the interaction between \textbf{User Index 10} and \textbf{Item Index 396} has been causing. The historical loss data initially indicated that the interaction was noisy, which was reflected in the initial confidence score of 0.916. However, subsequent actions to update the confidence scores and re-evaluate the model's performance have consistently resulted in worsened performance, suggesting that the interaction is not only noisy but also potentially malicious}.
  }
  \end{varwidth}}

\stitle{System Prompt of $Reflection^{R}$.}\vspace{3pt}\\
\setlength{\fboxsep}{5pt}
\colorbox{bgcolor}{
  \begin{varwidth}{\dimexpr\linewidth-2\fboxsep\relax}
  \small
  \emph{
    The current denoising rule is: \textcolor{purple}{<$M^{R}$>}. The current confidence scores and reasons are \textcolor{purple}{<$M^{A}$>}, and the reason for the agent's update denoising rule was: \textcolor{purple}{<$Reason^{A}$>}.
    Based on this information, follow these steps:\\
    1. Analyze current confidence scores and reasons, and identify which interactions are considered noisy and which are malicious, and summarize the reasoning for categorizing these interactions as noisy or malicious.\\
    2. Based on this analysis, compare the current denoising rule and update the denoising rule. Organize the rules hierarchically with descriptive classification labels to ensure they are clear, concise, and actionable, such as "Rule-1(Value-Related)", "Rule-1.1(Label)", "Rule-1.1.1(Label)", ..., "Rule-N(Label)", "Rule-N.1(Label)", etc., to organize different levels of rules.\\
    3. Merge similar rules, ensuring the rules are free of redundancy.\\
    4. Finally, output the updated denoising rules, and the response format should be strictly as follows:\\
    The updated denoising rules are: \color{RoyalBlue}{<New Rules>}.}
\end{varwidth}}

\stitle{Output Example of $Reflection^{R}$.}\vspace{3pt}\\
\setlength{\fboxsep}{5pt}
\colorbox{bgcolor}{
  \begin{varwidth}{\dimexpr\linewidth-2\fboxsep\relax}
  \small
  \emph{
    The updated denoising rules are:\\
    \textbf{Rule-1(Value-Related)}: The confidence of the interaction data is related to the loss value oft he interaction data, and the one with a large loss is more likely to be a noisy sample.\\
    \hspace*{1em}\textbf{- Rule-1.1(Value Threshold)}: The noisy sample's loss value exceeds the 95th percentile threshold.\\
    \hspace*{2em}\textbf{- Rule-1.1.1(Multiple Exceedances)}: If the loss of a sample exceeds the threshold three times across multiple trainings, it is classified as a noisy sample.\\
    \textbf{Rule-2(Fluctuation-Related)}: An interaction is considered noisy if the loss value shows a pattern of significant fluctuations over time.\\
    \hspace*{1em}\textbf{- Rule-2.1(High Variance)}: The loss value demonstrates high variance over consecutive time periods.\\
    \hspace*{2em}\textbf{- Rule-2.1.1(Variance Threshold)}: If the variance exceeds a threshold 0.5 within a specific number of time intervals, the sample is flagged as noisy.\\
    \hspace*{1em}\textbf{- Rule-2.2(Oscillation)}: The loss value alternates between extreme highs and lows without a consistent trend.\\
    \hspace*{2em}\textbf{- Rule-2.2.1(Oscillation Threshold)}: If the loss value oscillates beyond set upper bound (0.8) and lower bound (0.06) 4 times, it is marked as a noisy sample.\\
    \textbf{Rule-3(Outlier-Related)}: An interaction is considered noisy if the loss value is an outlier compared to the surrounding loss values in the dataset.\\
    \hspace*{1em}\textbf{- Rule-3.1(Median Comparison)}: The loss value is significantly higher or lower than the median loss of similar interactions.
  }
  \end{varwidth}}
\end{document}